\documentclass[preprint]{aastex631}

\received{June 5, 2024}
\revised{July 16, 2024}
\accepted{August 1, 2024}

\shorttitle{Cyclical period changes in Cataclysmic Variables}
\shortauthors{Souza \& Baptista}
\graphicspath{{./}{figures/}}

\begin{document}

\title{Cyclical period changes in cataclysmic variables: a statistical study}


 \correspondingauthor{Raymundo Baptista}
 \email{raybap@gmail.com}

 \author{Leandro Souza}
 \affiliation{Departamento de F\'{i}sica,
   Universidade Federal de Mato Grosso,
  Campus Cuiab\'a, Cuiab\'a, MT, Brazil}

 \author[0000-0001-5755-7000]{Raymundo Baptista}
 \affiliation{Departamento de F\'{i}sica, 
  Universidade Federal de Santa Catarina,
  Campus Trindade, Florian\'{o}polis, SC, Brazil}

 \begin{abstract}
   
We report the results of a statistical study of cyclical period changes
in cataclysmic variables (CVs). Assuming the third-body hypothesis as the
cause of period changes, we estimate the third-body mass, $m_3$, and its
separation from the binary, $a_3$, for 21 CVs showing cyclical period
changes from well-sampled observed-minus-calculated diagrams covering more
than a decade of observations. The inferred $a_3$ values are independent
of the binary orbital period, $P_\mathrm{orb}$, whereas the $m_3$ values
increase with $P_\mathrm{orb}$ by an order of magnitude from the
shortest period (oldest) to the longest period (youngest) systems,
implying significant mass loss from the third body with time.
A model for the time evolution of the triple system is not able to
simultaneously explain the observed behavior of the $m_3(P_\mathrm{orb})$
and $a_3(P_\mathrm{orb})$ distributions because the combined mass loss from
the binary and the third body demands an increase in orbital separation by
factors $\sim 140$ as the binary evolves toward shorter $P_\mathrm{orb}$'s,
in clear disagreement with the observed distribution. We conclude that
the third-body hypothesis is statistically inconsistent and cannot be
used to explain cyclical period changes observed in CVs. On the other
hand, the diagram of the amplitude of the period change versus the CV
donor-star mass is consistent both with the alternative hypothesis that
the observed cyclical period changes are a consequence of magnetic
activity in the solar-type donor star, and with the standard evolutionary
scenario for CVs.

\end{abstract}




\section{Introduction}

Cataclysmic variables (CVs) are compact binary systems where a solar-type,
lower main-sequence star (the secondary) overfills its Roche lobe and
transfers matter to a more massive white dwarf (WD). They correspond to
the advanced evolutionary stage of binaries with initial separations
$130 < a/R_{\odot} < 2000$ that passed through a common envelope phase (with
mass loss and significant loss of orbital angular momentum), leaving a
detached binary called a {\em precataclysmic}. The subsequent loss of
angular momentum further reduces the orbital separation until the secondary
star fills its Roche lobe and starts its mass-transfer phase as a CV
\citep{Warner,hellier}. In order to sustain mass transfer from the lower
mass star to the more massive WD, the binary must suffer continuous angular
momentum loss \citep[AML;][]{Warner,2011ApJS..194...28K}, either via
magnetic braking (MB) in the mass-donor stellar wind \citep[predominant for
long period systems, $P_\mathrm{orb}\geq 3\,h$; ][]{VerbuntZwaan81} or via
gravitational radiation \citep[GR, for short period systems, $P_\mathrm{orb}
\leq 2\,h$; ][]{Kraft62}, leading to a secular decrease of its orbital
period.

One interesting evolutionary question is whether the mass-transfer
process would lead to a long-term increase in WD mass, lending CVs as
possible progenitors of Type Ia supernovae \citep[via the accretion-induced
collapse of the WD to a neutron star, e.g.,][]{xraybin}.
However, detailed statistical studies \citep{1990LNP...369..392P,
  1998MNRAS.301..767S,2006MNRAS.373..484K,2011ApJS..194...28K}
show no evidence of an increase in WD mass along the evolution of the CV
binary, indicating that all mass transferred from the secondary star, and
temporarily accreted onto the WD, must be ejected from the system in
recurrent nova eruptions on timescales of $(10^4 - 10^5)$\,yr
\citep{1982ApJ...257..767F}. \citet{2006MNRAS.373..484K} finds an
average CV white dwarf mass of $\overline{M}_1 = 0.75 \pm 0.05\, M_{\odot}$.

The distribution of known CVs as a function of orbital period shows
a statistically significant dearth of objects in the period range
$2\,h \lesssim P_\mathrm{orb} \lesssim 3\,h$, known as the {\em period gap}
\citep{Warner,gansicke09}. Its identification prompted the idea that short-
and long-period systems could arise from two separate populations, with the
implication that long-period systems do not evolve across the gap.
Orbital evolution into and across the gap can be prevented either by
a `strong' MB (which would force long-period systems to bounce back to
longer periods at $P_\mathrm{orb} \simeq 3\,h$) or by an extremely
weak MB (such that no long-period system could have evolved across the gap
in a Hubble timescale). The first option predicts that secondaries at
$P_\mathrm{orb} \sim 3\,h$ must be degenerate stars with $M_2 < 0.08\,M_\odot$
and very depressed luminosities, while the second option requires very
low average secular mass-transfer rates above the gap ($\dot{M}_2
\lesssim 5\times 10^{-11}\,M_\odot yr^{-1}$). Both predictions are in marked
contrast to the observations, making the separate populations hypothesis
untenable \citep{Verbunt,Warner}.

The CV period gap is best explained in terms of the evolutionary
`disrupted-braking' model \citep[e.g.,][]{1991AA...248..525H}: the rapid
AML drives the mass-donor star out of thermal equilibrium and to be
increasingly oversized for its mass. A sudden reduction in MB efficiency
is believed to occur when the mass-donor star becomes fully convective at
the upper side of the period gap \citep[probably caused by the rearrangement
of its magnetic field structure with a reduction of open field lines;][]
{mt89}, allowing the bloated mass-donor star to detach from its Roche lobe,
halting mass-transfer, and to shrink back to its thermal-equilibrium radius.
Further AML (by GR and any residual MB) drives the now detached binary
across the period gap until mass transfer (and CV behavior) is resumed
when the donor star fills its Roche lobe again at the lower side of the
period gap.
There is significant observational support for the core hypothesis of
 orbital evolution across the period gap, i.e., that short- and long-period
 CVs comprise a unique population. First, the average WD mass is the same
 for long- and short-period systems \citep{1990LNP...369..392P,
  1998MNRAS.301..767S,2006MNRAS.373..484K,2011ApJS..194...28K}.
 Second, inferred donor star masses at the upper and lower end of the period
 gap nicely match; there are no donor stars with $M_2 \lesssim 0.2\,M_\odot$
 above the period gap and no donor stars with $M_2 \gtrsim 0.2\,M_\odot$
 below the period gap \citep{2011ApJS..194...28K}.
 Third, there is no period gap for magnetic systems \cite[possibly because
 in these cases the magnetic field lines of the mass-donor star are either
 closed or connected to the WD field lines, preventing outflows in the
 stellar wind, which powers the MB mechanism, thereby eliminating the
 cause of the period gap; e.g., ][]{1994.MNRAS.268.61,Warner}.

The fiducial mark in time provided by eclipses in CVs can usually be used
to determine $P_\mathrm{orb}$ (and its derivative) with high precision,
making it possible to detect and monitor variations in orbital period.
The reader is referred to \cite{2003MNRAS.345..889B} for a discussion
on how to minimize the influence of flickering and changes in bright spot
emission on the determination of mid-eclipse timings by averaging several
light curves to obtain a single, but robust mid-eclipse timing from a
sample of eclipse light curves, and is referred to \cite{2008AA...480..481B}
for a discussion on the influence of data sampling, time coverage, and
accuracy of eclipse timings on the ability to detect period modulations.
Accurate measurements of eclipse timings over an observational time span of
a few decades are required in order to distinguish between period variations
caused by AML \citep[steadily decreasing period;][]
{1983ApJ...275..713R,1988QJRAS..29....1K}, magnetic activity in the
solar-type star \citep[nonperiodic, cyclic variation;][]
{1992ApJ...385..621A,1999AA...349..887L,2006MNRAS.373..819L} or the
presence of a third body orbiting the binary \citep[strictly periodic
variation; e.g., ][]{2010AJ....140.1687Y}.
Remarkably, all CVs with well-sampled eclipse timings covering more than a
decade of observations show cyclical period changes on time scales of decades,
indicating that this is a widespread phenomenon requiring a common explanation
\citep{2003MNRAS.345..889B,2008AA...480..481B}.
The third-body hypothesis had already been discarded in the 1990s as a
plausible general explanation for the observed period changes
\citep[e.g,][]{1988Natur.336..129W,1992ApJ...385..621A,1994PASP..106.1075R,
  1999AA...349..887L}
because it implies a strictly periodic modulation, in contrast to the
nonperiodic or multiperiodic modulation observed in the objects with eclipse
timings measurements over the longest time span \citep[p.ex., UX\,UMa, RW\,Tri,
V2051\,Oph;][]{1991AJ....102.1176R,1995ApJ...448..395B,2003MNRAS.345..889B}.
It has, however, resurfaced in recent years in the wake of the wave of
discoveries of extrasolar planets\footnote{An updated list of discovered
  extrasolar planets can be found at https://exoplanet.eu/catalog/}
\citep[p.ex.;][]{barclay}, as the amplitudes of the modulation suggest
planetary masses for the resulting third bodies according to this hypothesis
\citep[e.g.,][]
{2011MNRAS.414L..16Q,2012MNRAS.425..749H,2012MNRAS.425..930G}.

In this paper, we test whether the third body hypothesis can provide a
consistent explanation for the overall observed cyclical period changes in
CVs when the secular evolution of these binaries is taken into account.
The data analysis is described in Sect.\,\ref{analise}; the discussion of
the results and the conclusions are presented in Sect.\,\ref{discuss}.

\section{ Data analysis } \label{analise}

We identified in the literature 21 CVs with well-sampled
observed-minus-calculated $(O-C)$ diagrams covering more than a decade of
eclipse timing observations. The observed cyclical period modulation timescale,
$P_\mathrm{mod}$, and amplitude, $A_\mathrm{mod}$, of the sample of objects
are listed in Table\,1, together with corresponding estimates of the orbital
inclination and masses of both stars. Given that the accuracy of mid-eclipse
timings is $\simeq$ (5--10)\,s  and $\simeq 20$\,s, respectively for short-
and long-period systems, and that the corresponding period modulation
amplitudes are in the ranges (20--60)\,s (short-period systems) and
(50--220)\,s (long-period systems), the statistical significance of
$A_\mathrm{mod}$ is typically above the 4$\sigma$ confidence level.
%
%
\begin{deluxetable*}{lccccccr}
\tablecaption{Relevant information for CVs with cyclical period changes}
\tablewidth{0pt}
\tablehead{
  \colhead{Object} & \colhead{$P_\mathrm{orb}$} & \colhead{$P_\mathrm{mod}$} &
  \colhead{$A_\mathrm{mod}$} & \colhead{$m_1/M_\odot$} & \colhead{$m_2/M_\odot$} &
  \colhead{$i(^o)$} & \colhead{References} \\
  & \colhead{(h)} & \colhead{(yr)} & \colhead{(s)}
}
\startdata
V4140 Sgr & 1.47 & $6.9 \pm 0.3$ & $17 \pm 3$ &
$0.73 \pm 0.08$   & $0.09 \pm 0.02$   & $80.2 \pm 0.5$ & 1,20 \\
DP Leo    & 1.49 & $28 \pm 2$  & $34 \pm 2$ & 
0.60              & 0.09              & 79.5           & 2,21 \\
V2051 Oph & 1.50 & $22 \pm 2$ & $17 \pm 3$ &
$0.78 \pm 0.06$   & $0.15 \pm 0.03$   & $83 \pm 2$     & 1,22 \\
OY Car    & 1.51 & $35.0 \pm 3.5$ & $46 \pm 3$ &
$0.685 \pm 0.011$ & $0.070 \pm 0.002$ & $83.3 \pm 0.2$ & 3,23 \\
EX Hya    & 1.64 & $17.5 \pm 1.2$ & $24 \pm 4$ &
$0.790 \pm 0.026$ & $0.108 \pm 0.008$ & $77.8 \pm 0.4$ & 4,24 \\
HT Cas    & 1.77 & $36 \pm 4$ & $40 \pm 5$ &
$0.61 \pm 0.04$   & $0.09 \pm 0.02$   & $81 \pm 1$     & 5,25 \\
Z Cha     & 1.78 & $28 \pm 2$ & $60 \pm 12$ &
$0.544 \pm 0.012$ & $0.081 \pm 0.003$ & $81.7 \pm 0.13$ & 6,26 \\
V893 Sco  & 1.82 & $10.2 \pm 0.22$ & $22 \pm 3$ &
0.89              & 0.175             & 72.5           & 7,27 \\
HU Aqr    & 2.08 & $11.96 \pm 1.41$ & $11 \pm 2$ &
$0.80 \pm 0.04$   & $0.18 \pm 0.06$   & $87.0 \pm 0.8$ & 8,28 \\
UZ For    & 2.10 & $23.4 \pm 5.1$ & $56 \pm 10$ &
0.7               & 0.14              & 80             & 9,29 \\
V348 Pup  & 2.44 & 14.7  & 34 &
0.65              & 0.20              & 80             & 10,30 \\
IP Peg    & 3.80 & $4.7 \pm 0.2$ & $94 \pm 8$ &
$1.09 \pm 0.10$   & $0.64 \pm 0.09$   & $79.3 \pm 0.9$ & 11,31 \\
UU Aqr    & 3.92 & 26 & 47 &
$0.67 \pm 0.14$   & $0.20 \pm 0.07$   & $78 \pm 2$     & 12,32 \\
U Gem     & 4.17 & 8 & 60 &
$1.20 \pm 0.05$   & $0.42 \pm 0.04$   & $69.7 \pm 0.7$ & 13,33 \\
DQ Her    & 4.64 & $17.7 \pm 0.3$ & 71 &
$0.60 \pm 0.07$   & $0.40 \pm 0.05$   & $86.5 \pm 1.6$ & 14,34 \\
UX UMa    & 4.72 & 30.4 & 81 &
$0.47 \pm 0.07$   & $0.47 \pm 0.10$   & $71.0 \pm 0.6$ & 15,35 \\
T Aur     & 4.91 & $23 \pm 2$ & $220 \pm 40$ &
0.68              & 0.63              & 57             & 5,36 \\
EX Dra    & 5.03 & 5 & 99  &
$0.75 \pm 0.15$   & $0.54 \pm 0.10$   & $85 \pm 3$     & 16,37 \\
RW Tri    & 5.57 & 13.6 & 73  &
0.7               & 0.6               & 75             & 17,38 \\
EM Cyg    & 6.98 & $17.74 \pm 0.01$ & $178 \pm 22$ &
$1.12 \pm 0.08$   & $0.99 \pm 0.12$   & $67 \pm 2$     & 18,39 \\
AC Cnc    & 7.21 & 16.2 & $225 \pm 17$ &
$0.76 \pm 0.03$   & $0.77 \pm 0.05$   & $75.6 \pm 0.7$ & 19,40 \\
\enddata
\tablecomments{References:
(1) \cite{2003MNRAS.345..889B}; (2) \cite{2011AA...526A..53B};
(3) \cite{2006MNRAS.372.1129G}; (4) \cite{1992IBVS.3724....1H};
(5) \cite{2008AA...480..481B}; (6) \cite{2002MNRAS.335L..75B};
(7) \cite{2014AA...566A.101B}; (8) \cite{2011MNRAS.414L..16Q};
(9) \cite{2010MNRAS.409.1195D}; (10) Borges \& Baptista (2020), priv. comm.;
(11) \cite{1993AA...273..160W}; (12) \cite{2014RMxAC..44S.148B};
(13) \cite{1988Natur.336..129W}; (14) \cite{2009AA...503..883D};
(15) \cite{1991PASP..103.1258R}; (16) \cite{2003PASP..115.1105S};
(17) \cite{1991AJ....102.1176R}; (18) \cite{2010PASJ...62..965D};
(19) \cite{2007AA...466..589Q}; (20) \cite{2005ASPC..330..365B};
(21) \cite{2002AA...392..541S}; (22) \cite{1998MNRAS.300..233B};
(23) \cite{1989ApJ...341..974W}; (24) \cite{2008AA...480..199B};
(25) \cite{1991ApJ...378..271H}; (26) \cite{1986MNRAS.219..629W};
(27) \cite{2001ApJ...563..351M}; (28) \cite{2009AA...496..833S};
(29) \cite{2011MNRAS.416.2202P}; (30) \cite{2001MNRAS.328..903R};
(31) \cite{1988MNRAS.231.1117M}; (32) \cite{1994ApJ...433..332B};
(33) \cite{2007AJ....134..262E}; (34) \cite{1993ApJ...410..357H};
(35) \cite{1995ApJ...448..395B}; (36) \cite{1980MNRAS.192..127B};
(37) \cite{2003PASP..115.1105S}; (38) \cite{2004AA...417..283G};
(39) \cite{2000MNRAS.313..383N}; (40) \cite{2004MNRAS.353.1135T}.
}

\end{deluxetable*}

First-generation planets around binaries are formed from the same
protostellar disk as the binary itself and are expected to be approximately
coplanar to the binary plane \citep{1994MNRAS.269L..45B}. Therefore, we
assume that the inclination of the orbit of the third body is the binary
inclination. Given that the inclination enters the equations only through
 $\sin i$ and in most cases $i \geq 70^o$, small deviations from this
assumption are not expected to lead to significant errors in the results.
We note that this assumption does not hold if the triple system is the result
of a gravitational capture of a planet-size body by the binary. In this
case, there is no correlation between the binary plane and the triple system
orbital plane. However, while it is arguable that a planet would form from
the protostellar disk in {\em every} CV, it seems much more unlikely that a
planet-size body be gravitationally captured in {\em every single} CV
(as the widespread occurrence of cyclical period changes implies).

Assuming that the observed period modulation is the consequence of the
light-travel time effect caused by the wobbling of the CV binary around the
center of mass of a triple system, and that the third body orbital plane
is co-planar to that of the CV binary, we use the data of Table\,1 to
calculate the mass of the third body, $m_3$, and its separation to the
CV binary, $a_3$. Since $M_{12} = M_1+M_2$ is the mass of the binary,
the mass of the third body is given by
\begin{equation}
  m_{3} = \frac{A_\mathrm{mod}\,c\,(2 \pi)^{2/3} (M_{12} + m_{3})^{2/3}}
  {G^{1/3} P_\mathrm{mod}^{2/3} \sin i} \, ,
  \label{eq-mass}
\end{equation}
and the separation between the binary and the third body is given by
\begin{equation}
  a_{3} = G^{1/3}\,(M_{12} + m_{3})^{1/3}
  \left(\frac{P_\mathrm{mod}}{2 \pi}\right)^{2/3} \, ,
\end{equation}
where $G$ is the gravitational constant. Note that since $m_3$ appears
on both sides of Eq.\,(\ref{eq-mass}), it has to be solved iteratively.

\begin{figure}
  \begin{center}
\includegraphics[scale=0.75]{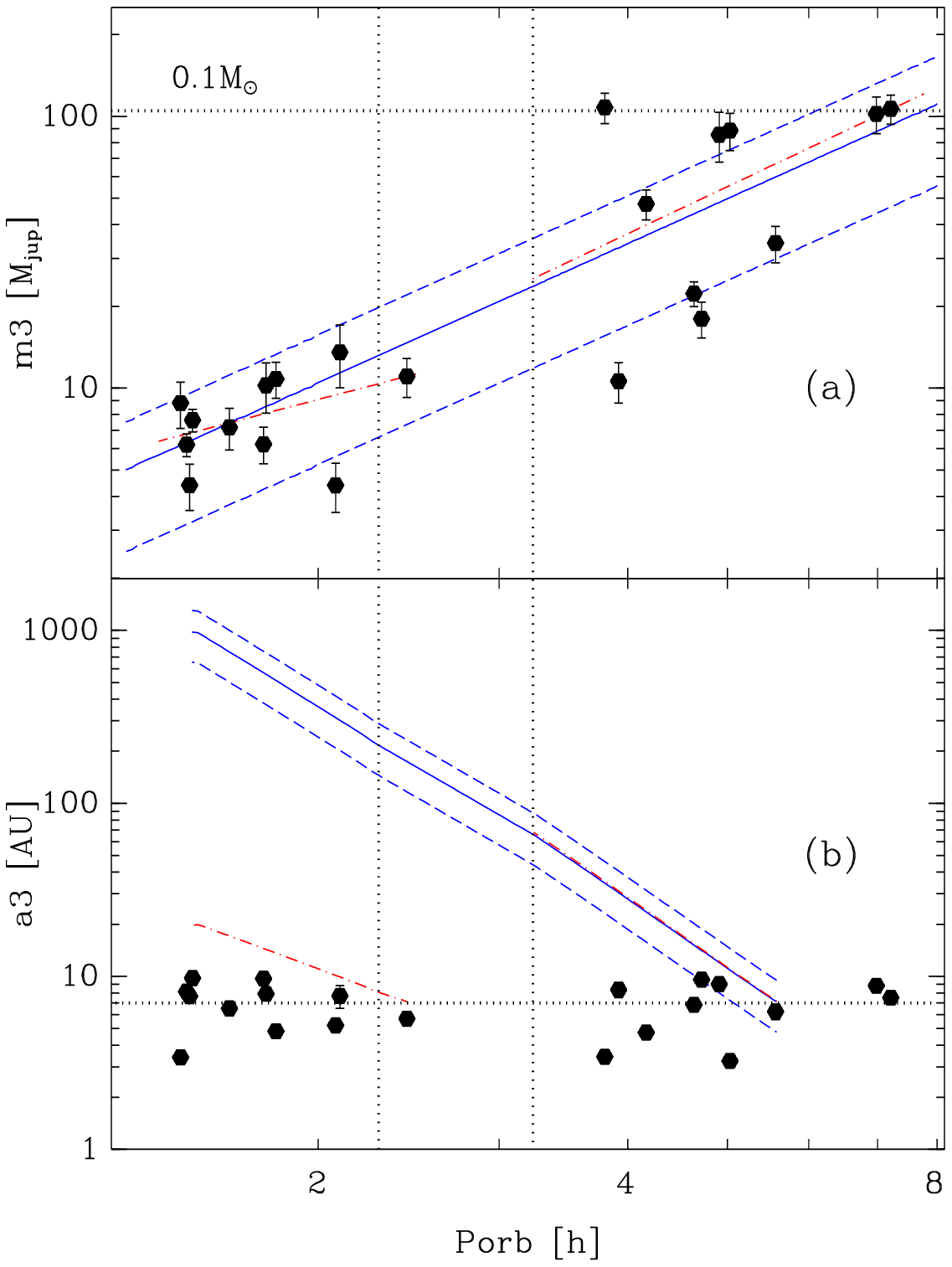}
\caption{(a) The distribution $m_3(P_\mathrm{orb})$. Vertical dotted lines
  indicate the period gap; an horizontal dotted line depicts the value
  $m_3 = 0.1\,M_\odot$. Blue lines show the best-fit empirical model
  $\log m_3(t) = A + B\log P_\mathrm{orb}(t)$ (solid line) and its
  uncertainties at the 1$\sigma$ level (dashed lines). Red dotted-dashed
  lines show separate fits to the short- and long-period systems.
  (b) The corresponding $a_3(P_\mathrm{orb})$ distribution. The horizontal
  dotted line indicate the average value $\overline{a}_3= 7\,au$. The blue
  lines show the $a_3$ values predicted by the evolutionary model of the
  triple system (solid line) and its uncertainties at the 1$\sigma$ level
  (dashed lines), for an initial value $a_3(P_\mathrm{orb}=5.6\,h)= 7\,au$.
  Red dotted-dashed lines show the $a_3(P_\mathrm{orb})$ distributions
  corresponding to the separate $m_3(P_\mathrm{orb})$ fits to the short-
  and long-period systems.
  \label{fig1}}
\end{center}
\end{figure}

The $m_3$ and $a_3$ values obtained under the third-body hypothesis are
plotted in the upper and lower panels of Figure\,\ref{fig1} as a function
of $P_\mathrm{orb}$. The uncertainties in $m_3$ and $a_3$ were obtained by
randomly varying the values of $P_\mathrm{mod}$, $A_\mathrm{mod}$, $M_1$,
$M_2$, and $i$ assuming gaussian distributions of standard deviations
equal to their respective uncertainties with a Monte Carlo code using
$10^5$ trials. The listed $P_\mathrm{mod}$ errors for IP\,Peg and EM\,Cyg
are unrealistically small and do not take into account uncertainties
arising from gaps of years without observations of eclipse timings in
their $(O-C)$ diagrams; we adopt a conservative stance and assume a formal
relative error of 10\% percent in both cases. We also assume a formal
relative error of 10\% for all other quantities for which no uncertainty
is provided.

The derived $m_3$ values increase with orbital period, by 1 order of
magnitude from the shortest period systems (i.e., the oldest CVs) to
the longest period systems (i.e., the youngest CVs).
Given that CVs evolve toward shorter orbital periods, the third-body
hypothesis implies in systematic and significant mass loss from the third
body with time. We may express this secular mass loss from the third body
with an empirical model,
\begin{equation}
  \log m_3(t) = A + B \log P_\mathrm{orb}(t) \, ,
  \label{eq:perda}
\end{equation}
where $A= 0.46 \pm 0.13$ and $B= 1.77 \pm 0.23$ are the best-fit
coefficients by least squares.  In addition, $m_3$ values for the longer
periods imply masses large enough to characterize these bodies not as
planets, but as low-mass stars that are expected to contribute to the
system spectrum (particularly in the infrared) -- there is currently
no observational support for this prediction.

On the other hand,
the inferred $a_3$ values are independent of $P_\mathrm{orb}$ and are in
the range $3 \leq a_3(au) \leq 10$ ($700 \leq a_3/R_\odot \leq 2300$),
with an average value of $\overline{a}_3 = 7\,au$.
We note that these values are comparable to the initial separation of the
companion binary, implying that either the third-body hypothesis is
inconsistent because the resulting triple system would be gravitationally
unstable (with the third body being either accreted or ejected from the
system in a timescale much shorter than the binary lifespan) or that the
inferred $a_3$ has shrunk from a much larger initial orbital separation
(possibly as a consequence of AML during the common envelope phase).

In a triple system consisting of a close binary gravitationally coupled to
a distant third body, the total mass of the triple system is
\begin{equation}
M_t = M_{12} + m_3 \, ,
\end{equation}
while its total angular momentum is
\begin{equation}
J_3 = M_{12}\, m_3 \left(\frac{G\, a_3}{M_t}\right)^{1/2} \, .
\end{equation}
Applying a logarithmic time derivative to the above equation, we obtain
\begin{equation}
  \frac{\dot{J_3}}{J_3} = \frac{\dot{M}_{12}}{M_{12}} +\frac{\dot{m}_3}{m_3} +
  \frac{1}{2} \frac{\dot{a}_3}{a_3} - \frac{1}{2} \frac{\dot{M_t}}{ M_t}\, .  
\end{equation}
We assume that the fractional variation of the total mass is approximately
equal to the fractional variation of the binary mass (i.e., $M_t \gg m_3$),
and that the total angular momentum of the triple system is conserved
during the CV binary lifespan, $\dot{J_3}=0$, to find
\begin{equation}
  \frac{ \dot{a}_3}{a_3} =
  -2\,\frac{\dot{m}_3}{m_3} - \frac{\dot{M}_{12}}{M_{12}} \, .
  \label{eq-evolve}
\end{equation}
Because $\dot{m}_3<0$ (empirical third-body mass loss) and $\dot{M}_{12}<0$
(in recurrent nova eruptions), it is unavoidable to have $\dot{a}_3>0$,
i.e., the triple system separation must increase with time, at a rate
which is more significant the more pronounced are the mass losses
$\dot{m_3}$ and $\dot{M}_{12}$.

We developed a computer model to trace the evolution of the orbital
separation of the triple system, $\dot{a_3}/a_3$, by combining a
semiempirical evolutionary model for mass and AML
from the CV binary \citep{2011ApJS..194...28K}, which provides the
$\dot{M}_{12}/M_{12}$ term, with the empirical inferred mass loss from the
third body of Eq.\,(\ref{eq:perda}), which yields the $2\,\dot{m_3}/m_3$
term.  The outcome of the model is a theoretical $a_3(P_\mathrm{orb})$
distribution, which can then be compared to the observed distribution.

The semiempirical CV donor sequence and evolutionary track of
  \cite{2011ApJS..194...28K} starts at $P_\mathrm{orb}= 5.6\,h$ (because
  above this limit the secondaries are expected to be evolved stars).
  They adopt a power-law mass-radius relation of the type $R_2 \propto M_2^\xi$,
  with best-fit values of $\xi=0.61$ below the period gap and $\xi=0.69$
  above the period gap, and assume that AMLs are driven by GR and by the MB
  prescription of \cite{1983ApJ...275..713R}, respectively, below and above
  the period gap, with best-fit scaling factors of $f_\mathrm{GR}= 2.47$ and
  $f_\mathrm{MB}= 0.66$.
It is also assumed that all matter transferred to the WD ends up being
ejected from the system in consecutive nova eruptions and, therefore, WD
mass remains constant throughout the evolution ($\dot{M}_1= 0, \dot{M}_{12}=
\dot{M}_2$) at $M_1= 0.75\,M_\odot$. Given that the time interval between
successive nova eruptions ($10^4 - 10^5$\,yr) is much smaller than the
evolutionary timescale ($10^8 - 10^9$\,yr) and the amount of mass transferred
($\Delta M \sim 10^{-5} - 10^{-4}\, M_\odot$) is negligible in comparison to
the WD mass, the process is smooth and indistinguishable from that of a
continuous mass loss from the binary.
The Roche-lobe-filling secondary stars of CVs obey a well-known
period--density relationship \citep[e.g.,][]{Warner},
\begin{equation}
  \overline{\rho}=
  \frac{3 M_2}{4 \pi R_2^3}\simeq \frac{107}{P_\mathrm{orb}^2}\, g\, cm^{-3} \, .
  \label{eq:period-density}
\end{equation}
Applying a logarithmic time derivative to Eq.(\ref{eq:period-density})
and combining the result with the mass--radius relation, we obtain the
expression for the evolution of the orbital period of the binary,
\begin{equation}
  \frac{\dot{P}_\mathrm{orb}}{P_\mathrm{orb}}=
  \frac{3\xi - 1}{2} \frac{\dot{M}_2}{M_2} \, ,
\end{equation}
indicating that CVs must evolve towards shorter periods
($\dot{P}_\mathrm{orb}<0$) provided $\dot{M}_2<0$ and $\xi > 1/3$.

The code to compute the evolution of the orbital separation of the
triple system starts at an orbital period
$P_\mathrm{orb}=5.6\,h$ with an initial separation $a_3= 7\,au$.
At each iteration, the $m_3$ value is computed from Eq.\,(\ref{eq:perda}),
the binary parameters $M_2, \dot{M}_{2}$, and $\dot{P}_\mathrm{orb}$ are
taken from the CV evolutionary model and the parameters of the triple
system ($ M_1, M_2, P_\mathrm{orb}, m_3$ and $a_3$) are updated. Mass
transfer (and CV binary mass loss) is consistently interrupted while
the binary evolves across the period gap; during this phase, the
$a_3(P_\mathrm{orb}$) distribution is driven uniquely by the
$2\,\dot{m_3}/m_3$ term.

Applying a logarithmic time derivative to Eq.(\ref{eq:perda}) we obtain
\begin{equation}
  \frac{2\,\dot{m}_3}{m_3}= 2\,B \frac{\dot{P}_\mathrm{orb}}{P_\mathrm{orb}}
  = B (3\xi - 1) \frac{\dot{M}_2}{M_2} \, .
  \label{eq:triple}
\end{equation}
Noting that
\begin{equation}
  \frac{\dot{M}_{12}}{M_{12}}= \frac{M_2}{M_1+M_2} \frac{\dot{M}_2}{M_2} \, ,
  \label{eq:binary}
\end{equation}
we combine Eqs.(\ref{eq:triple}-\ref{eq:binary}) to evaluate the relative
importance of the $2\,\dot{m_3}/m_3$ and $\dot{M}_{12}/M_{12}$ terms to the
evolution of the orbital separation of the triple system,
\begin{equation}
  \eta= \frac{2 \dot{m_3}/m_3}{\dot{M}_{12}/M_{12}}=
  B (3\xi - 1) \left( \frac{M_1+M_2}{M_2} \right) \, .
\end{equation}
$\eta$ varies from $\simeq 4$ (at $P_\mathrm{orb}= 5.6\,h$, with $\xi=0.69$)
to $\simeq 13.7$ (at $P_\mathrm{orb}= 1.5\,h$, with $\xi=0.61$), indicating
that the slope of the theoretical $a_3(P_\mathrm{orb})$ distribution is
largely dominated by the contribution from the $2\,\dot{m_3}/m_3$ term.
This can be seen by the modest decrease in the slope of the distribution
at the period gap, where the contribution from the $\dot{M_2}/M_2$ term
vanishes (Figure\,\ref{fig1}).  It also indicates that the details of the
CV binary evolution have little effect in the resulting theoretical
$a_3(P_\mathrm{orb})$ distribution, particularly at the low $P_\mathrm{orb}$
end. We confirmed that by testing different common prescriptions for the
secondary star mass-radius relation and for the MB law \cite[e.g.,][]
{Warner,2011ApJS..194...28K}. The resulting differences are small in
comparison to the uncertainties in the theoretical $a_3(P_\mathrm{orb})$
distribution and do not alter the following conclusions.

The lower panel of Fig.\,\ref{fig1} compares the observed $a_3$ values
with those predicted by the theoretical model taking into account the
inferred mass loss from the third body. The inconsistency of the
third-body hypothesis is clear; the discrepancy cannot be eliminated by
assuming different fits to the $m_3(P_\mathrm{orb})$ distribution nor
by considering different models of CV binary evolution.  The central
point is that the combination of the drastic third-body mass reduction
with time (by a factor $\simeq 20$) and the additional evolutionary
CV binary mass loss implies in an increase in orbital separation
$a_3$ by factors $\sim 140$, in marked contrast to the observations.

In order to flatten the theoretical $a_3(P_\mathrm{orb})$ distribution to
match the observations, one needs to postulate a very significant AML
from the triple system, $\dot{J_3}$, throughout the CV binary lifespan,
and it would need to be very fine-tuned in order to properly compensate
the contributions of the $2\,\dot{m_3}/m_3$ and $\dot{M_2}/M_2$ terms and
result in an observed $a_3$ distribution independent of $P_\mathrm{orb}$.
Such AML cannot be accounted for by either gravitational
radiation (because it is negligible at such large orbital separations),
magnetic braking (as there is no tidal coupling mechanism to connect
the AML of the magnetic mass-donor star in the CV
binary to the far distant third body), or
frictional angular momentum losses \citep[FAML; e.g.,][]{MacDonald86}
from the repeated interaction of the third body with a nova ejecta
(as the scaling of the FAML effect to the mass, size, and separation of
the third body indicates that the resulting $\dot{J}_3/J_3$ term is
$\sim 2-5$ orders of magnitude smaller than the $2\,\dot{m_3}/m_3$ and
$\dot{M_2}/M_2$ terms).

In a last-ditch attempt to save the third-body hypothesis, we put
aside the assumption that long-period systems evolve across the period
gap and we treat the short- and long-period systems as separate
populations. Separate fits to the short- and long-period system
$m_3(P_\mathrm{orb})$ distributions are shown as red dotted-dashed lines
in the upper panel of Fig.\,\ref{fig1}. We find $B_s= 0.99 \pm 0.39$
(statistical significance of $2.5\sigma$) and $B_l= 1.79 \pm 0.28$
(statistical significance of $6.4\sigma$), respectively, for the
short- and long-period systems, indicating that the inferred $m_3$
values increase with $P_\mathrm{orb}$ in both cases. Because CVs on both
groups still evolve toward shorter periods via AML ($\dot{M_2}<0$,
$\dot{P}_\mathrm{orb}<0$), the results continue to imply in third-body
mass loss with time and, consequently, in a prediction of a significant
increase in triple system separation with time. The corresponding
$a_3(P_\mathrm{orb})$ distributions are shown as red dotted-dashed lines
in the lower panel of Fig.\,\ref{fig1}. The discrepancy remains.
The third-body hipothesis is inconsistent even if we drop the core
assumption of CV evolution across the period gap.

\section{Discussion and Conclusions} \label{discuss}

We use a statistical approach to explore the implications of the hypothesis
that cyclical period changes observed in CVs are a consequence of the
presence of a circumbinary body modulating the mid-eclipse times. We use
the equation of light-travel time to the observer to calculate values of
the third-body mass $m_3$ and orbital separation $a_3$ from the amplitude
and time scale of the cyclical period variation for a set of 21 CVs with
well-sampled $(O-C)$ diagrams covering more than a decade of observations.
The derived $m_3$ values increase with orbital period; $m_3$ values for
the shortest period systems are 1 order of magnitude smaller than
for the longest period systems. On the other hand, the derived $a_3$
values are independent of $P_\mathrm{orb}$ ($\overline{a}_3 = 7$ au).

In an attempt to explain these results, we developed a model that
considers both the mass and AML losses from the CV binary as well as the
implied mass loss from the third body. The model cannot simultaneously
explain the behavior of the $m_3 (P_\mathrm{orb})$ and $a_3 (P_\mathrm{orb})$
distributions. The basic problem is that the combined mass loss of the
binary and the third body imposes a significant increase in the orbital
separation $a_3$ (by factors $\sim 140$) as the CV binary evolves toward
smaller $P_\mathrm{orb}$'s, in clear disagreement with the observed
distribution.

We conclude that the third-body hypothesis is statistically inconsistent
and cannot be used to explain the observed widespread cyclical period
changes in CVs. Although it is possible to model the orbital period
modulation in individual systems with the circumbinary third-body hypothesis,
most cases are discarded by additional data \citep{2014MNRAS.445.1924B}
or by detailed analysis of orbital dynamic stability
\citep{2012AJ....144...34H,2012MNRAS.425..749H,2013MNRAS.431.2150W}.

\begin{figure}
  \begin{center}
\includegraphics[scale=0.7,angle=270]{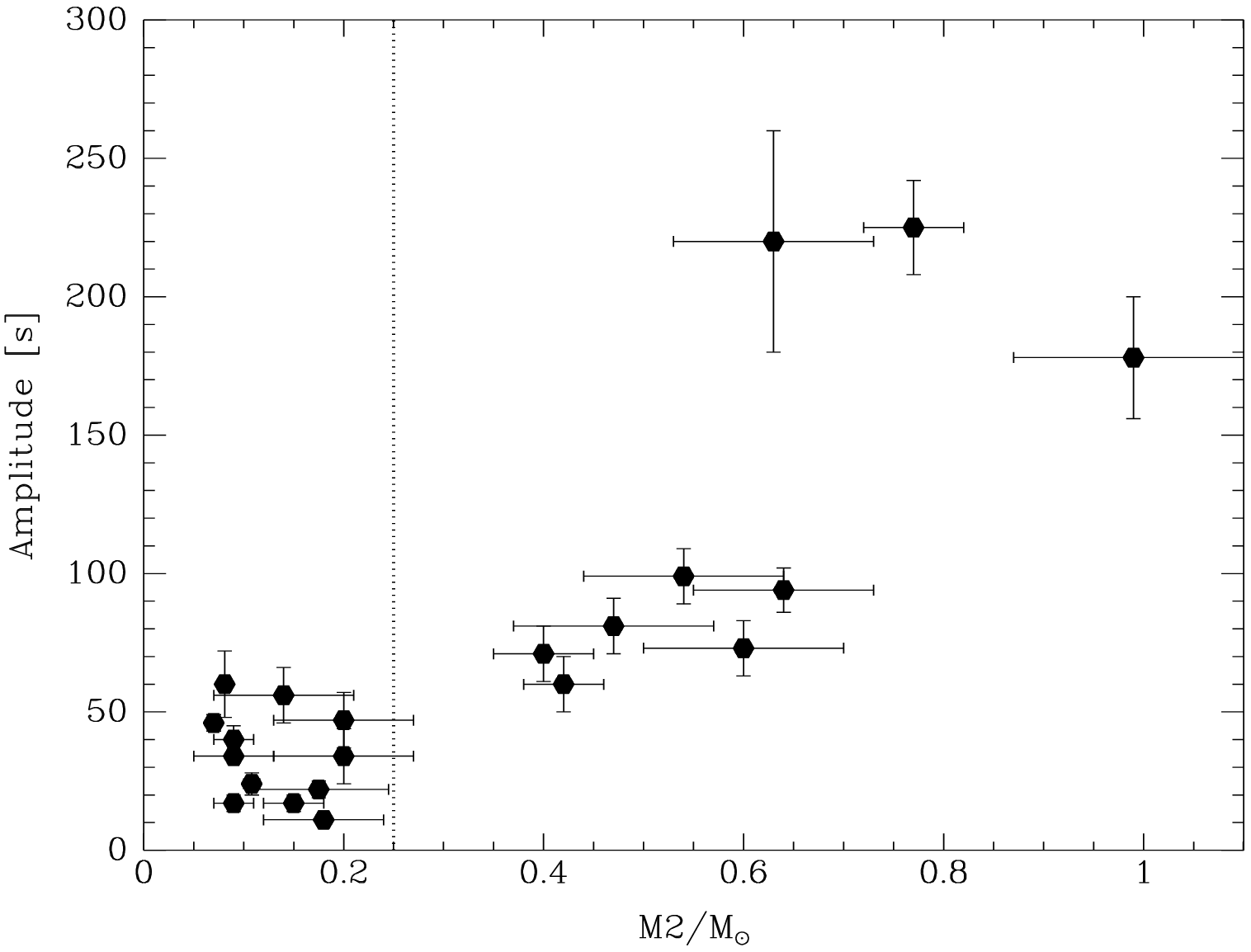}
\caption{Distribution $A_{mod} \times M_2$. A vertical dotted line marks
  the $M_2= 0.25\,M_\odot$ limit where mass-donor stars are expected to
  become fully convective. The amplitude of the period modulation
  increases with $M_2$ for secondary stars with convective envelopes and
  is systematically lower and flat for fully convective secondary stars.
  \label{fig2}}
\end{center}
\end{figure}

The remaining hypothesis is that the observed cyclical period changes are
a consequence of the magnetic activity in the mass-donor star
\citep[e.g,][]{1994PASP..106.1075R,2006MNRAS.373..819L}, a phenomenon
probably present in all CVs -- in accordance with the widespread
occurrence of cyclical period changes.
A consequence of the Applegate/Lanza hypothesis is the prediction that
there should be a change in behavior and a reduction in the amplitude of
the period modulation in systems with fully convective secondaries, in
connection with the explanation for the period gap. Figure\,\ref{fig2}
shows the distribution $A_\mathrm{mod} \times M_2$. The observed correlation
between $A_\mathrm{mod}$ and $M_2$ in the mass range where the secondaries
have convective envelopes and the flattening of the distribution together
with lower $A_\mathrm{mod}$ values in the mass range where the secondaries
are fully convective are consistent with the above prediction
and in line with the common explanation for the period gap.
%

\begin{acknowledgments}
  We thank the anonymous referee for useful comments and suggestions that
  helped to improve the presentation of our results.
  L.S. acknowledges financial support from CAPES/Brazil.
\end{acknowledgments}


\end{document}